\documentclass[aip,preprint]{revtex4-1}
\usepackage{graphicx}
\usepackage{amssymb}
\usepackage{color}
\usepackage{amsbsy}
\usepackage[intlimits]{amsmath}
\usepackage{amsthm}
\usepackage{amsfonts}
\usepackage{natbib}
\begin{document}
\title{Pinning-induced \emph{pn} junction formation in low-bandgap two-dimensional semiconducting systems}
\author{David A. Deen} \email{david.deen@alumni.nd.edu}
\affiliation{independent submission}
\begin{abstract}
A model is presented for $pn$ junction formation near metal-semiconductor contacts in two-dimensional semiconducting systems such as graphene. Carrier type switching occurs in a region near the metal-semiconductor junction when energy band bending leads to a crossing between the junction Fermi level and the Dirac energy. A bias-dependent depletion region occurs due to the minimization of carrier density, which is shown to act as an additional parasitic resistance in devices. The $pn$ junction resistance is demonstrated by its implementation in a transfer length structure.
\end{abstract}
\maketitle
The advent of two-dimensional (2D) semiconducting systems, such as graphene and the transition-metal dichalcogenides, has driven significant efforts to utilize these systems for electronic devices.\cite{Geim,MoS2} In the vast majority of embodiments, electronic devices like the field effect transistor and various sensor designs utilize a modulated channel.\cite{Champlain} In those cases, metallic electrodes are used to make ohmic contact to the semiconducting channel. Most studies address the minimization of parasitic access resistance via the contact resistance ($R_c$) of the contacts to graphene.\cite{Chaves,Robinson,Moon} However, little attention has been given to the manifestation of spuriously high-resistance $pn$ junctions formed in close proximity to the contact. The formation of the $pn$ junction is due to the contrast between the Fermi-energy pinning at the metal-semiconductor interface and opposite carrier typing outside the contact region that results from ambient electrostatic bias. Ultimately, the $pn$ junction resistance results as an additive parasitic resistance to the other parasitic resistances (access, $R_{acc}$, and $R_c$) present, which may subsequently serve to impair device performance. This work presents the development of an electrostatic model that describes the formation of such a $pn$ junction with low carrier density in proximity to a metal-2D semiconductor contact. The energy intersection between the pinned Fermi level under the metal-semiconductor contact and the 2D semiconductor Dirac energy ($E_D$) is central to the development and results in the low carrier density bias-dependent $pn$ junction. The model includes calculations for the width of minimum conductivity, its dependence on voltage bias, and how it is manifested in a conventional test structure. More complex effects such as Klein tunneling and the role of band chirality on carrier generation-recombination are not incorporated in the electrostatic model but likely play a role in the transport picture. \cite{Klein}

\begin{figure}
\includegraphics[width=\columnwidth]{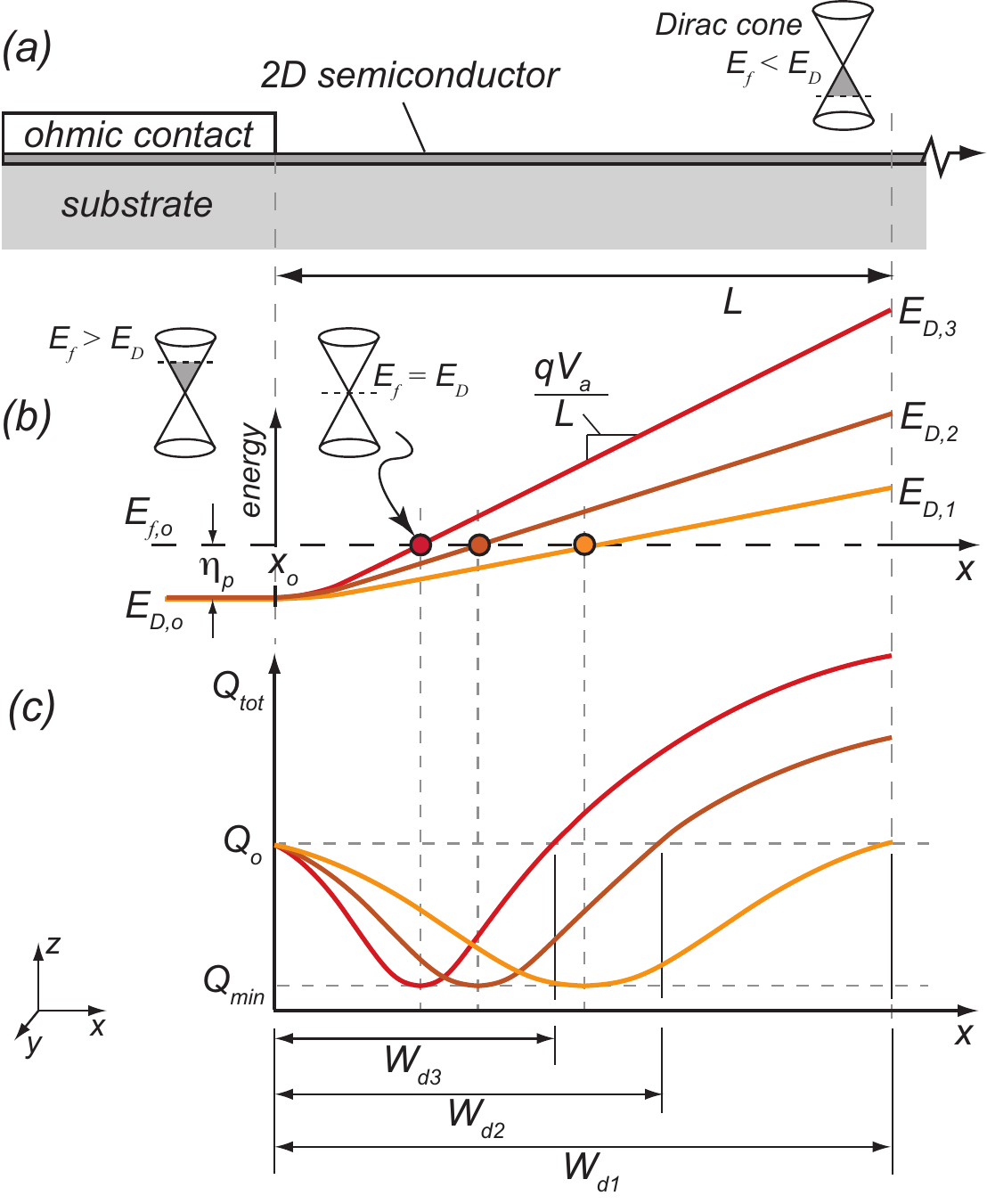}
\caption{(a) Illustration of the 2D structure, (b) lateral band diagram in graphene along the source-drain direction at several different voltage biases, and (c) the resultant carrier density profiles along the graphene channel.}
\label{fig:one}
\end{figure}

Graphene and some transition-metal dichalcogenide semiconductors exhibit a unique linear energy dispersion relation due to hybridized sp$^2$ covalent bonding within their bravais lattice. In this work zero-bandgap graphene is used as an example from the broad 2D semiconducting families. Figure \ref{fig:one} shows an illustration of a graphene device cross-section near one ohmic contact (a) with the linearized energy band diagram (b) and resultant total charge distribution (c) for several voltage biases. The ohmic contact was assumed to pin the graphene Fermi level at an energy of $\eta_p$ only under the contact.\cite{Chaves,Robinson,Giovannetti} The linearization of the energy band diagram is an approximation made only in a small energy range about the Fermi level-Dirac energy (F-D) intersection where total charge density is minimized and resistance has a maximum. Realistically, the Dirac energy (or conduction and valence band intersection) bends throughout the length of the device in the $x$-direction in coherence with a change in charge density (Poisson). However, for the small energy range about the F-D crossing the linearized energy is sufficient to describe the carrier typing and therefore, the electrostatics of the system in the vicinity of the ohmic contact.  

An approximation has been made for the resistance associated with the region beyond the linear energy region where $|(E_D - E_f)| \gg |\eta_p|$. In this region, the total charge density exponentially increases according to Fermi-Dirac statistics and yields a resistance much less than in the region where the F-D crossing occurs. Therefore, this low-resistance extension has been treated as the sheet resistance outside the proximity of the F-D intersection (regions beyond $W_d$ in Fig. 1 (c)). The pinning energy polarity at the metal-graphene junction and the ambient electrostatic bias away from the contact are responsible for setting up the conditions leading to the F-D intersection and therefore, the attributes of the $pn$ junction. Those conditions determined the polarity of the $E_D$ slope, number of F-D intersection within a given region, and length of total charge minimum referred to as the `depletion width' ($W_d$) in this work. If two F-D intersections occur, the total resistance becomes approximately twice the resistance of a single $pn$ junction.

The carrier distribution in graphene results from the linear dispersion relation and therefore follows from the full Fermi-Dirac integral, $n,p(\eta) = qN_G\mathcal{F}_i(\pm \eta)$ where $\mathcal{F}_i(\pm \eta) = (\Gamma(i+1))^{-1} \int_0^{\infty} u^i(1+e^{\mp \eta}e^u)^{-1} du$.\cite{Champlain, Jena} The longitudinally-dependent total charge is given as
\begin{equation}\label{eq:one}
Q_{tot}(x) = \frac{qN_G}{\Gamma(2)} \left[\int_0^{\infty} \frac{u du}{1+e^{u}e^{\phi_F(x)}} - \int_0^{\infty} \frac{u du}{1+e^{u}e^{-\phi_F(x)}}\right]
\end{equation}
where $N_G = (g_sg_v/2\pi)(k_BT/\hbar v_F)^2$ is the 2D density of states with $g_s$ and $g_v$ are the spin and valley degeneracies, respectively, $v_F$ is the Fermi velocity, $q$ the elemental charge, $\Gamma(\dots)$ is the gamma function, $\hbar$ is the reduced Planck's constant, $\phi_F(x) = k_BT \eta_F(x)$ is the energy along the channel normalized by the thermal energy $k_BT$, $k_B$ is Boltzmann's constant, and $T$ the temperature. The linearized energy band was approximated as, $E - E_f = V_{a}(x-x_o)/L$, where $x_o$ is at the ohmic contact edge as illustrated in Fig. \ref{fig:one}. The normalized longitudinal energy term used in Eq. \ref{eq:one} follows from the boundary conditions of the linear energy band,
\begin{equation}
\eta_F(x) = \frac{qV_{a}}{L}x - \eta_p 
\end{equation}
where $L$ is the length of the graphene sheet between the two biased contacts as defined in Fig. \ref{fig:one}, $V_{a}$ is the applied voltage, and $\eta_p$ is the contact pinning energy. The depletion width is defined as twice the distance from contact pinning to the Fermi level crossing and follows
\begin{equation}\label{eq:width}
W_d=\frac{2\eta_p L}{q V_{a}}
\end{equation}
In the conventional case of a three-dimensional semiconductor $pn$ junction, the depletion width is defined by the balance of the internal field and the diffusion of free carriers. The linear energy approximation accounts for the complete electrostatic picture for the pinning-induced $pn$ junction in graphene. The total resistance of the depletion width accounting for the single Fermi level crossing, is given by the integral relationship,
\begin{equation}\label{eq:R}
R_d = \frac{1}{W_g}\int_0^{W_d} \frac{dx}{\sigma (x)} = \frac{1}{qW_g} \int_0^{W_d} \frac{dx}{\mu_n n(x)+\mu_p p(x)}
\end{equation}
where $\sigma (x) = q(\mu_nn(x)+\mu_pp(x))$ is the graphene conductivity and is dependent on the areal charge densities, $n$ and $p$, and the respective carrier mobilities, $\mu_n$ and $\mu_p$, and $W_g$ is the cross-sectional width of the graphene sheet. 

\begin{figure}
\includegraphics[width=\columnwidth]{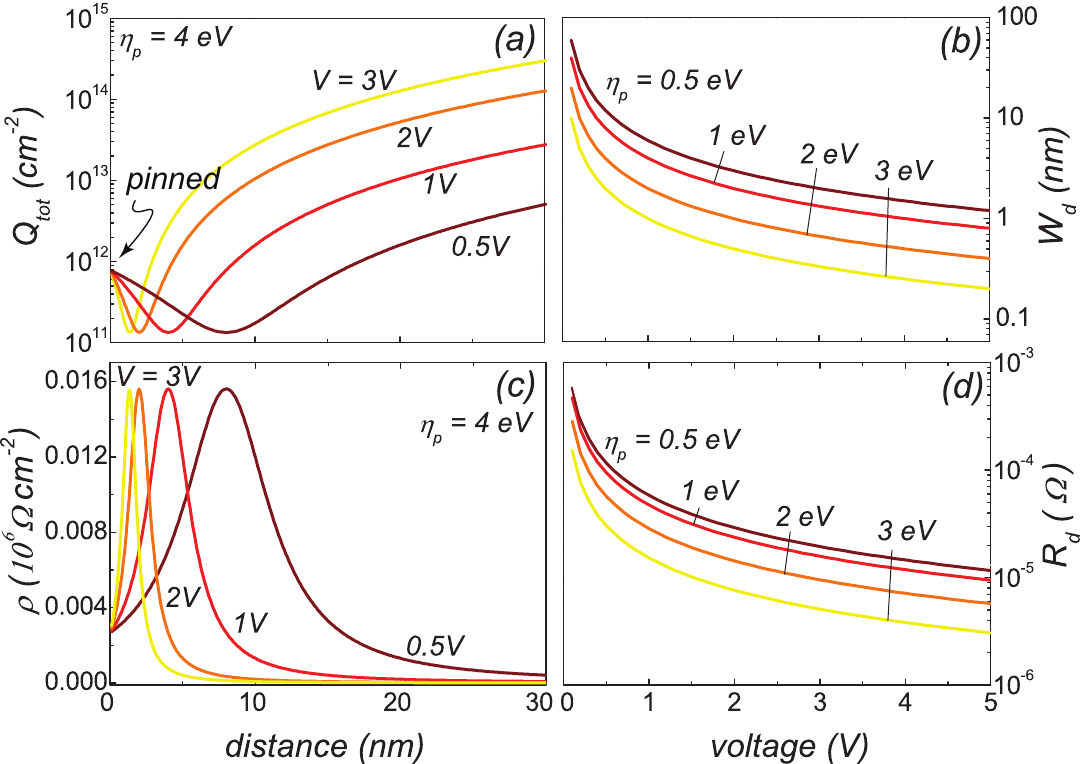}
\caption{Parametric characteristics for the $pn$ junctions formed in graphene in proximity to an ohmic contact; (a) Charge density versus longitudinal distance, (b) resistivity versus distance for various applied voltages, (c) depletion width versus applied longitudinal voltage, and (d) depletion region resistance versus applied voltage for various pinning values.}
\label{fig:two}
\end{figure}
Parametric calculations are shown in Fig. \ref{fig:two} wherein Figs. \ref{fig:two} (a) and (c) show total charge density and correspondent resistivity ($\rho(x) = 1/\sigma(x)$) in the graphene sheet, respectively, as a function of distance away from the ohmic contact for several longitudinal voltages. A pinning energy of 4 eV has been used, which is commensurate with values reported. \cite{Chaves,Robinson,Moon} The charge density and resistivity minima progressively decreased with an increase in applied voltage as the F-D intersection migrated toward the ohmic contact. The rapidly diminished value of $\rho$ away from the F-D crossing (total charge minima) validates the previously stated approximation. Figure \ref{fig:two} (b) and (d) show depletion width and depletion region resistance in the graphene sheet, respectively, as a function of applied voltage for several values of pinning energy. As a greater bias voltage is applied, energy-band bending is enhanced and the slope of the linear region becomes greater leading to the trends shown in Fig. \ref{fig:two}. The bias-dependence of $R_d$ is of note within the context that parasitic resistances are typically fixed (e.g. $R_c$ or $R_{acc}$). A dynamic parasitic resistance like $R_d$ may serve to complicate the extraction of other intrinsic device metrics (i.e. transistor transconductance, frequency performance, etc.). One method that may prove to disable the formation of the $pn$ junction is doping graphene. By doping, the Fermi level in graphene can be significantly shifted such that no F-D intersection may occur. Both $W_d$ and $R_d$ linearly decreased with increasing voltage following Eq. \ref{eq:width} and \ref{eq:R}, respectively. 

\begin{figure}
\includegraphics[width=\columnwidth]{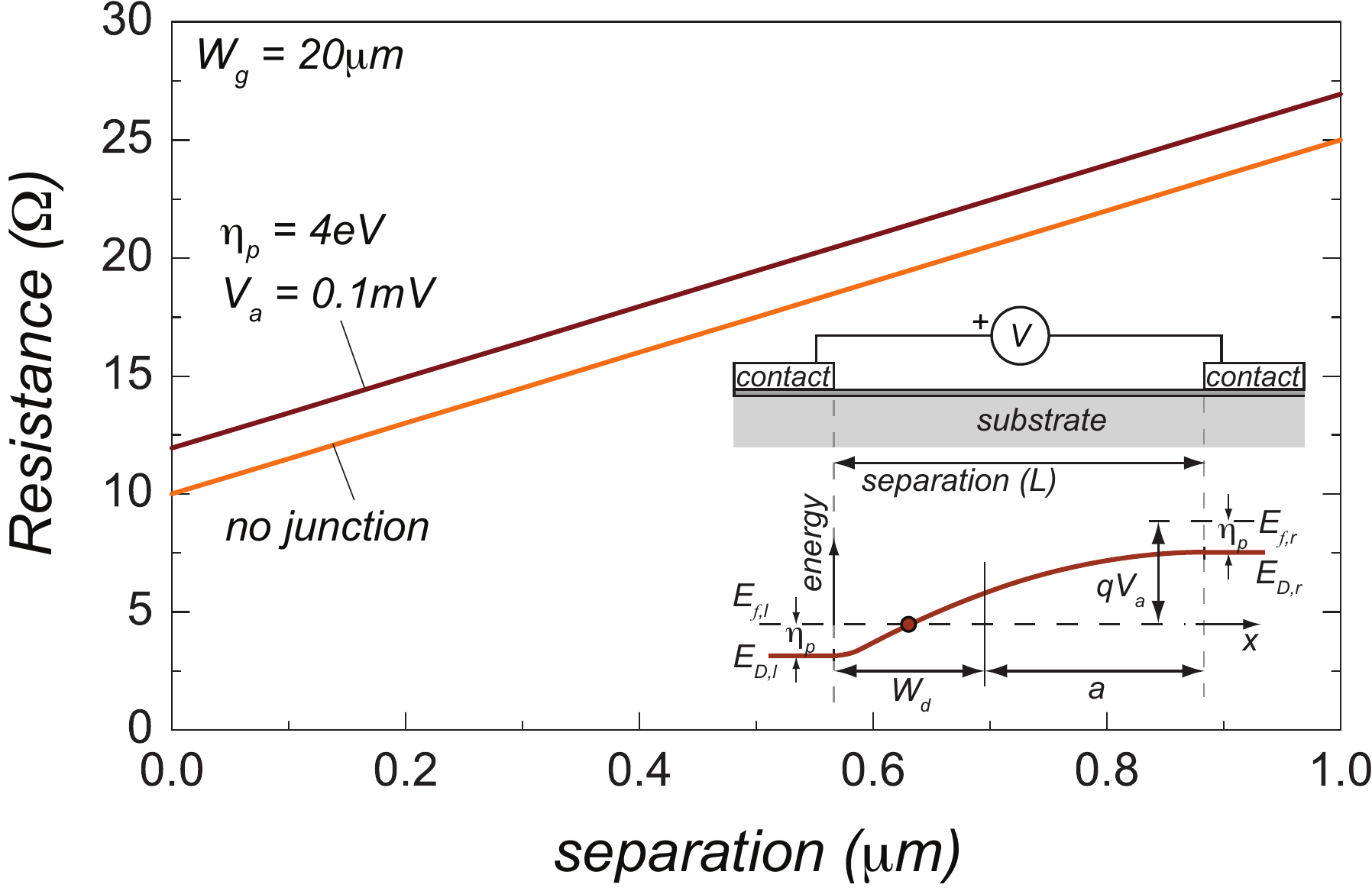}
\caption{Resistance of TLM structure comprised by graphene with and without the inclusion of a single $pn$ junction showing a corresponding resistance increase. The inset shows the graphene TLM configuration and 1D linearized energy band diagram between two contacts.}
\label{fig:three}
\end{figure}
The effect of the $pn$ junction on a device is illustrated through implementation into a transfer length structure (TLM). The TLM structure is a conventional test structure to extract information on $R_c$ and $R_{sh}$. The total resistance of a TLM structure is conventionally given by $R(L) = 2R_c + \frac{L}{W}R_{sh}$, where $R_{sh}$ is the sheet resistance. For the scenario considered here, we assume there is a single $pn$ junction between TLM contacts and the depletion width is small compared to the separation between contacts, $W_d \ll L$. The latter assumption is verified by the calculation of $W_d$ as shown in Fig. \ref{fig:two}(b). These assumptions result in the TLM resistance relationship for graphene,
\begin{equation}
R_{TLM} = 2R_c + R_d + R_{sh}\left(\frac{L-a}{W_g}\right)
\end{equation}
with $R_d$ being the depletion region resistance as given by Eq. \ref{eq:R} and $a$ is the distance between the contacts and the depletion width as illustrated in the inset in Fig. \ref{fig:three}. Written in terms of the longitudinal graphene conductivity and invoking the constraint, $L \gg a$, the total resistance of a graphene TLM structure with a single $pn$ junction follows,
\begin{equation}\label{eq:RTLM}
R(L) = 2R_c + \frac{L}{W_g}R_{sh} + \underbrace{\frac{1}{W_g} \int_0^{W_d} \frac{dx}{\sigma(x)}}_{R_d}
\end{equation}
Equation \ref{eq:RTLM} shows that the presence of a single $pn$ junction in the graphene sheet adds a resistance proportional to the total bias-dependent resistivity of the junction to the TLM resistance. The $pn$ junction resistance is independent of separation, $L$, and therefore is additive. The manifestation of additional $pn$ junction resistance is demonstrated in Fig. \ref{fig:three} where a single $pn$ junction is included in a TLM resistance curve with a Fermi pinning of 4 eV compared to a curve that does not include a pinning-induced $pn$ junction. Other pertinent graphene parameters used in the calculation are $W_g$ = 20 $\mu m$, $R_{sh}$ = 300 $\Omega/\Box$, $\mu_n = \mu_p$ = 1000 $cm^2/Vs$, $T$ = 300K, and $v_F$ = $1.1\times10^6$ $cm/s$. The linear energy approximation should hold for other 2D low bandgap semiconductors.

An electrostatic model has been presented for $pn$ junction formation in graphene in close proximity to ohmic contacts where Fermi level pinning occurs. It has been shown that electrostatic depletion of charge within the $pn$ junction introduces a bias-dependent high-resistance region. The effect of pinning-induced $pn$ junction formation in graphene has been illustrated for the case of a TLM structure, wherein the junction resistance was shown to be an additive term to the total TLM resistance. The broader impact of $pn$ junction formation in low-bandgap two-dimensional systems may be spurious resistance that impairs device performance.

We would like to thank J. G. Champlain at the Naval Research Laboratory for helpful technical discussion.

\end{document}